\documentclass[fleqn,usenatbib]{mnras}
\usepackage{newtxtext,newtxmath}
\usepackage[T1]{fontenc}
\usepackage{ae,aecompl}
\usepackage{cancel}
\usepackage{graphicx}	
\usepackage{amsmath}	
\usepackage{amssymb}	
\usepackage[inline]{enumitem}

\title[Paper]{Deep multi-survey classification of variable stars}

\author[Carlos Aguirre et al.]{
C. Aguirre,$^{1}$\thanks{E-mail: claguirre@uc.cl }
K. Pichara,$^{1,2,3}$ \thanks{E-mail: kpb@ing.puc.cl }
I. Becker$^{1}$ \thanks{E-mail: iebecker@uc.cl }
\\
$^{1}$Computer Science Department, Pontificia Universidad Cat\'{o}lica de Chile.\\
$^{2}$Millennium Institute of Astrophysics, Chile\\
$^{3}$Institute of Applied Computational Science (IACS),
Harvard. Cambridge, MA, USA.
}

\date{Accepted XXX. Received YYY; in original form ZZZ}

\pubyear{2017}

\begin{document}
\label{firstpage}
\pagerange{\pageref{firstpage}--\pageref{lastpage}}
\maketitle

\begin{abstract}
  During the last decade, a considerable amount of effort has been made to classify variable stars using different machine learning techniques. Typically, light curves are represented as vectors of statistical descriptors or features that are used to train various algorithms. These features demand big computational powers that can last from hours to days, making impossible to create scalable and efficient ways of automatically classifying variable stars. Also, light curves from different surveys cannot be integrated and analyzed together when using features, 
because of observational differences. For example, having variations in cadence and filters, feature distributions become biased and require expensive data-calibration models. The vast amount of data that will be generated soon make necessary to develop scalable machine learning architectures without expensive integration techniques. 
   Convolutional Neural Networks have shown impressing results in raw image classification and representation within the machine learning literature. 
In this work, we present a novel Deep Learning model for light curve classification, mainly based on convolutional units. Our architecture receives as input the differences between time and magnitude of light curves. It captures the essential classification patterns regardless of cadence and filter. 
In addition, we introduce a novel data augmentation schema for unevenly sampled time series. 
We tested our method using three different surveys: OGLE-III; Corot; and VVV, which differ in filters, cadence, and area of the sky. We show that besides the benefit of scalability, our model obtains state of the art levels accuracy in light curve classification benchmarks. 
 \end{abstract}



\section{Introduction} 
    \label{sec:intro}
    There has been a considerable amount of effort trying to automate the classification of variable stars \citep{debosscher2007automated,sarro2009automated,bloom2011data,pichara2012improved, nun2015fats, mackenzie2016clustering, pichara2016meta, benavente2017automatic}. Variable stars such as RR Lyrae, Mira, and Cepheids are important for distant ladder measurements as shown in \citealt{bloom2011data}. The ability to classify variable stars is closely related to the way light curves are represented. One way is to create vectors of statistical descriptors, called features, to represent each light curve \citep{bloom2011data, nun2015fats}. One of the most popular set of statistical features is presented in \citet{nun2015fats}, also known as FATS features (stands for Feature Analysis for Time Series). These vectors demand large computational resources and aim to represent the most relevant characteristics of light curves. A lot of effort has been made to design these features, and the creation of new ones implies a lot of time and research. The future surveys in Astronomy demand new ways of extracting these features. One example of the huge amount of data is The Large Synoptic Survey Telescope (LSST) \citep{borne2007data,abell2009lsst} that will start operating on 2022. It is estimated that the LSST will produce 15 to 30TB of data per night. New ways of treating this information have been proposed \citep{mackenzie2016clustering,valenzuela2017unsupervised, naul2017recurrent, gieseke2017convolutional}. \citealt{mackenzie2016clustering}'s uses an unsupervised feature learning algorithm to classify variable stars. \citealt{valenzuela2017unsupervised} perform unsupervised classification by extracting local patterns among light curves and create a ranking of similarity. To extract such patterns they use a sliding window as done in \citealt{mackenzie2016clustering}.
    
\begin{figure*}    \includegraphics[width=\textwidth,]{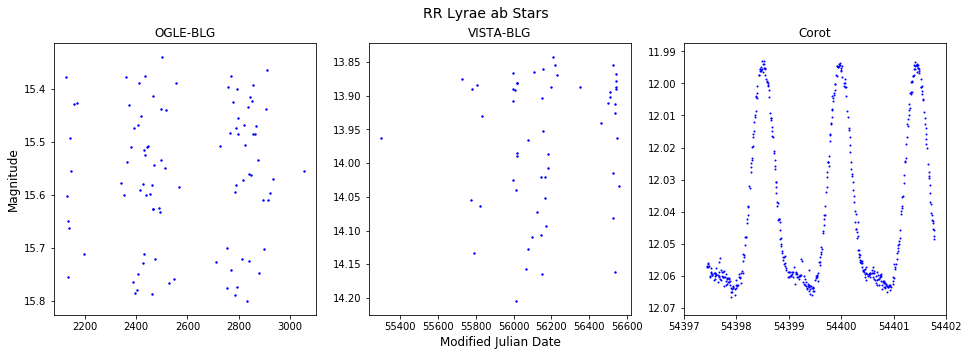}
  \includegraphics[width=\textwidth,]{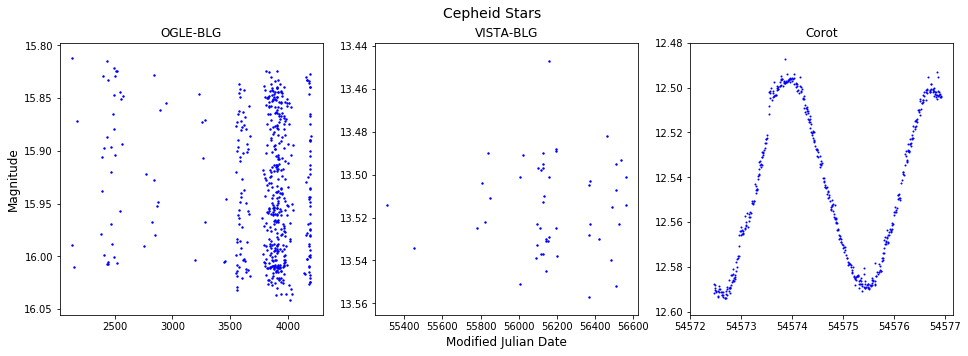}
     \caption{Comparison of RR Lyrae ab and Cepheid stars in OGLE-III, VISTA and Corot Survey respectively. Difference between magnitude and cadence is shown.}
    \label{fig:light_curves_comparison}
\end{figure*}
    
\begin{figure}    \includegraphics[width=\columnwidth,]{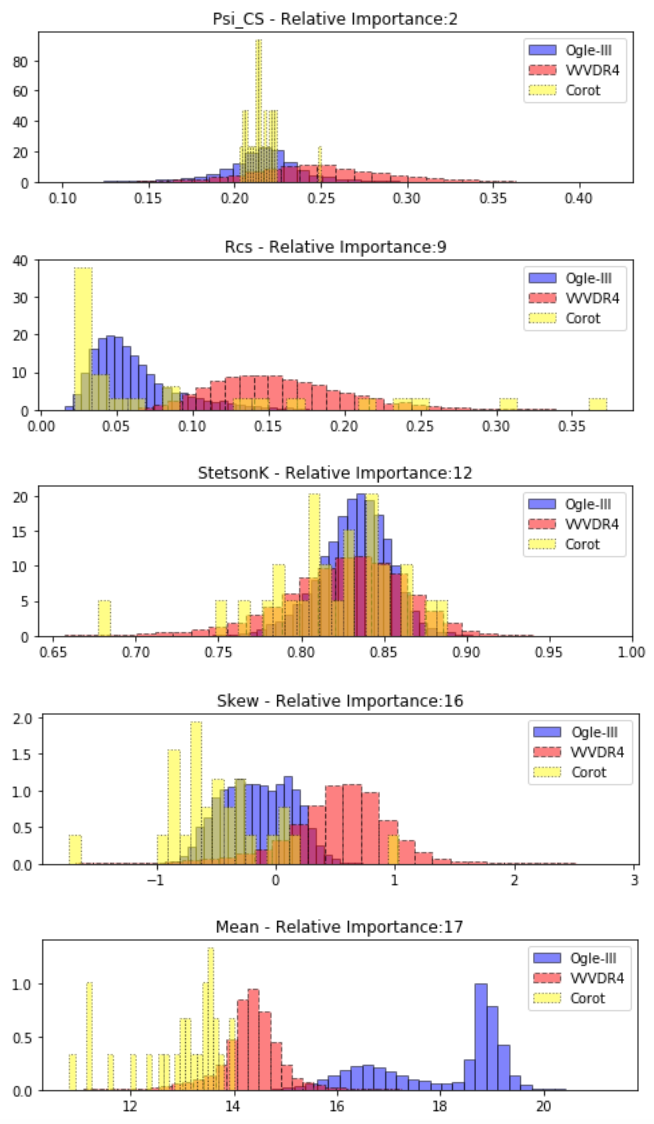}
    \caption{Comparison of FATS feature in RR Lyrae ab using histogram plots of stars using OGLE-III, Corot and Vista Surveys. Every feature is shown with its relative importance in classification as mention in \citealt{nun2015fats}.}
    \label{fig:fats_comparison}
\end{figure}

     Either way, none of these methods can be applied immediately in new surveys. The limited overlap and depth coverage among surveys makes difficult to share data. The difference in filter and cadence makes it even harder without any transformation to the data. Figure \ref{fig:light_curves_comparison} shows an example of the complexity that exists among stars and surveys. Light curves have a difference in magnitude and time, and most of the time they are not human-eye recognizable, even by experts. Since all the magnitudes are calibrated by using statistics, it does not work correctly because of underlying differences between surveys. Figure \ref{fig:fats_comparison} shows a comparison of statistical features of RR Lyrae ab stars using three different catalogs. To the best of our knowledge, little efforts have been made to create invariant training sets within the datasets. \citealt{benavente2017automatic} proposed an automatic survey invariant model of variable stars transforming FATS statistical vectors \citep{nun2015fats} from one survey to another. As previously mentioned, these features have the problem of being computationally expensive and the creation of new ones implies a lot of time and research. Therefore, there is a necessity of faster techniques able to use data from different surveys.
    
    Artificial neural networks (ANNs) have been known for decades \citep{cybenko1989approximation,hornik1991approximation}, but the vast amount of data needed to train them made them infeasible in the past. The power of current telescopes and the amount of data they generate have practically solved the problem. The improves in technology and the big amount of data makes ANNs feasible for the future challenges in astronomy. 
    
    Artificial neural networks or deep neural networks create their own representation by combining and encoding the input data using non-linear functions \citep{lecun2015deep}. Depending on the number of hidden layers, the capacity of extracting features improve, together with the need for more data \citep{lecun2015deep}. Convolutional neural networks (CNNs) are a particular type of neural network that have shown essential advantages in extracting features from images \citep{krizhevsky2012imagenet}. CNNs use filters and convolutions that respond to patterns of different spatial frequency, allowing the network to learn how to capture the most critical underlying patterns and beat most of the classification challenges \citep{krizhevsky2012imagenet}. Time series, as well as in images, have also proven to be a suitable field for CNNs \citep{zheng2014time,jiang2016cryptocurrency}. 
 
    In this paper, we propose a convolutional neural network architecture that uses raw light curves from different surveys. Our model can encode light curves and classify between classes and subclasses of variability. Our approach does not calculate any of the statistical features such as the ones proposed in FATS, making our model scalable to vast amounts of data. In addition, we present a novel data augmentation schema, specific for light curves, used to balance the proportion of training data among different classes of variability. Data augmentation techniques are widely used in the image community \citep{krizhevsky2012imagenet,dieleman2015rotation,gieseke2017convolutional}.
    
      We present an experimental analysis using three datasets: OGLE-III \citep{udalski2004optical}, VISTA \citep{minniti2010vista} and Corot \citep{baglin2002corot,borde2003exoplanet}. The used datasets differ in filters, cadence and observed sky-area. Our approach obtains comparative results with a Random Forest (RF) classifier, the most used model for light curve classification \citep{richards2011machine,dubath2011random,long2012optimizing,gieseke2017convolutional}, that uses statistical features. Finally, we produce a catalog of variable sources by cross-matching VISTA and OGLE-III and made it available for the community. \footnote{Datasets will be available in http://gawa.academy.org/profile/<authorusername>/. A Data Warehouse for astronomy \citep{Machin2018}}.     
    
    The remainder of this paper is organized as follows: Section \ref{sec:related} gives an account of previous work on variable stars classification and convolutional neural networks. In Section \ref{sec:background} we introduce relevant background theory. Section \ref{sec:method} explains our architecture and Section \ref{sec:data} describes the datasets that are used in our experiments. Section \ref{sec:data_extraction} explains the modifications that are made to the data and Section \ref{sec:par_initialization} gives an account of the parameters that are used in the architecture. Section \ref{sec:results} shows our experimental analysis and Section \ref{sec:time_analysis} presents a study of the time complexity of our model. Finally, in Section \ref{sec:conclusion}, the main conclusions and future work directions are given.

\section{Related Work}
    \label{sec:related}
    As mention before, there has been huge efforts to classify variable stars \citep{richards2011machine, nun2015fats, nun2014supervised, huijse2014computational, mackenzie2016clustering, pichara:2016,Valenzuela:2017}. The main approach has been the extraction of features that represent the information of  light curves. \citealt{debosscher2007automated} was the first one proposing 28 different features extracted from the photometric analysis. \citealt{sarro2009automated} continue the work by introducing the color information using the OGLE survey and carrying out an extensive error analysis.
\citealt{richards2011machine} use 32 periodic features as well as kurtosis, skewness, standard deviation, and stetson, among others, for variable stars classification. \citealt{pichara2012improved} improve quasars detection by using a boosted tree ensembles with continuous auto regressive features (CAR). \citealt{pichara2013automatic} introduce a probabilistic graphical model to classify variable stars by using catalogs with missing data.


\citealt{kim2014epoch} use 22 features for classifying classes and subclasses of variable stars using random forest. \citealt{nun2015fats} published a library that facilitates the extraction of features of light curves named FATS (Feature Analysis for Time Series). More than 65 features are compiled and put together in a Python library\footnote{Information about the features and manuals of how to use them are available as well.}. \citealt{kim2016package} publish a library for variable stars classification among seven classes and subclasses. The library extracts sixteen features that are considered survey-invariant and uses random forest for doing the classification process. 

  A novel approach that differs from most of the previous papers is proposed by \citealt{mackenzie2016clustering}. He face the light curve representation problem by designing and implementing an unsupervised feature learning algorithm. His work uses a sliding window that moves over the light curve and get most of the underlying patterns that represent every light curve. This window extracts features that are as good as traditional statistics, solving the problem of high computational power and removing the human from the pre-processing loop. \citealt{mackenzie2016clustering} work shows that automatically learning the representation of light curves is possible.
    
    Since every survey has different cadences and filters, research has been done to learn how to transfer information among them. The ability to use labeled information without an extensive work accelerates the process of bringing new labeled datasets to the community. \citealt{long2012optimizing} propose using noise and periodicity to match distributions of features between catalogs. Also, show that light curves with the same source and different surveys would normally have different values for their features. \citealt{benavente2017automatic} represent the joint distribution of a subset of FATS features from two surveys and creates a transformation between them by using a probabilistic graphical model with approximate inference. Even though he use FATS for his experiments, they can be easily changed to \citealt{mackenzie2016clustering} features, making the process even faster. \citealt{pichara2016meta} present a meta-classification approach for variable stars. He show an interesting framework that combines classifiers previously trained in different sets of classes and features. His approach avoids to re-train from scratch on new classification problems. The algorithm learns the meta-classifier by using features, but including their computational cost in the optimization of the model.

    All methods mentioned above invest lots of efforts finding a way to represent light curves. There have been several works in deep learning where the network itself is the one in charge of learning the representation of data needed for classification. \citealt{baglin2002corot} use a vanilla neural network for classifying microlensing light curves from other types of curves such as variable stars. \citealt{belokurov2003light} continue the work presenting two neural networks for microlensing detection. \citealt{krizhevsky2012imagenet} show the importance of CNNs for image-feature-extraction, and use them to classify, achieving impressive results. \citealt{zeiler2014visualizing} study the importance of using filters in each convolutional layer and explain the feature extraction using the Imagenet dataset. \citealt{dieleman2015rotation} apply galaxy morphology classification using deep convolutional neural networks. \citealt{cabrera2017deep} use a rotation-invariant convolutional neural network to classify transients stars in the HITS survey. \citealt{mahabal2017deep} transform light curves into a two-dimensional array and perform classification with a convolutional neural network.

	CNNs not only work on images. Many studies have been done using one-dimensional time series. \citealt{zheng2014time} use them to classify patient's heartbeat using a multi-band convolutional neural network on the electrocardiograph time series. \citealt{jiang2016cryptocurrency} create a decision maker for a cryptocurrency portfolio using a CNN on the daily price information of each coin. 
   
\section{Background Theory}
    \label{sec:background}
    
    In this section, we introduce the basics on both artificial neural networks and convolutional neural networks, to gain the necessary insights on how our method works.
 
    \subsection{Artificial Neural Networks}
    Artificial neural networks (ANN) are computational models based on the structure and functions of biological axons \citep{basheer2000artificial}. The ability for axons to transmit, learn and forget information has been the inspiration for neural networks. ANNs are capable of extracting complex patterns from the input data using nonlinear functions and learn from a vast amount of observed data. ANNs have been used in different areas such as speech recognition \citep{graves2013speech, xiong2016achieving,xiong2017microsoft}, image recognition \citep{krizhevsky2012imagenet, szegedy2015going, ren2015faster}, and language translation \citep{jean2014using, sutskever2014sequence} among others.
  
    The basic forming unit of a neural network is the perceptron (as axon in biology). As shown in Figure \ref{fig:perceptron} a perceptron takes inputs and combine them producing one output. For each of the input, it has an associative weight $w_i$ that represent the importance of that input, and a bias $b_0$ is added to each perceptron. The perceptron combines those inputs with their respective weights in a linear form and then uses a non-linear activation function to produce the output:

\begin{figure}    \includegraphics[width=\columnwidth,]{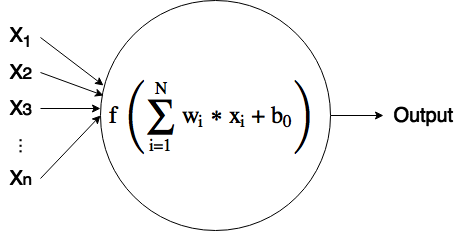}
    \caption{The forming basic unit the perceptron. It combines the inputs with their respective weights and apply a non-linear function to produce the output. A bias $b_0$ is added to each perceptron letting it change the starting point.}
    \label{fig:perceptron}
\end{figure} 

\begin{equation}
    Output = f\left( \sum_{i=1}^{N} w_i \cdot x_i  + b_0 \right )
\end{equation}

 Where $f$ is the activation function. Two of the most widely used activation functions are the $tanh$ and the $sigmoid$ function because of their space complexity. Moreover, $relu$ functions are widely used in convolutions as they avoid saturation and are less computationally expensive.
 
\begin{equation}
sigmoid(x) \quad = \quad \frac{1}{1+ e^{-x}}
\end{equation}
\begin{equation}
tan(x) \quad = \quad \frac{e^{x} - e^{-x}}{e^{x} + e^{-x}}
\end{equation}
\begin{equation}
relu(x) \quad = \quad max(0,\:x)
\end{equation}
    
\begin{figure}    \includegraphics[width=\columnwidth]{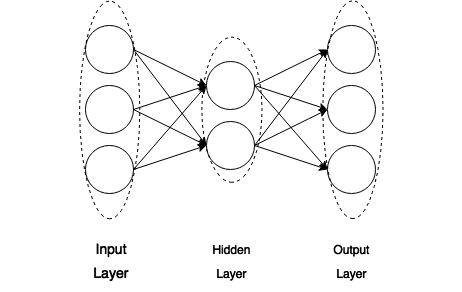}
    \caption{ A vanilla neural network with an input, hidden and output layer. Total number of hyperparameters 17.}
    \label{fig:vanilla}
\end{figure}

    The most basic neural network is the vanilla architecture consisting of three layers: \begin{enumerate*} \item the input layer, \item the hidden layer and \item the output layer.\end{enumerate*} As shown in Figure \ref{fig:vanilla} the input of a perceptron is the output of the previous one, except for the input layer that does not have any input and the output layer that does not have any output. A fully connected layer is when every neuron in one layer connects to every neuron in the other one. The vanilla architecture consists of two fully connected layers.
    
 The number of perceptrons for each layer depends on the architecture chosen and therefore the complexity of the model. A neural network can have hundreds, thousands or millions of them. The experience of the team, as well as experimenting different architectures, is critical for choosing the number of layers, perceptrons for each one and the number of filters to be used. The number of hyperparameters is mainly given by the weights in the architecture. The input layer is where we submit our data and has as many neurons as our input does. The hidden layer is the one in charge of combining the inputs and creating a suitable representation. Finally, the number of neurons in the output layer is as many classes we want to classify. 

    Many architectures have been proposed for artificial neural networks. The vanilla architecture can be modified in the number of hidden layers and the number of perceptrons per layer. ANNs with one hidden layer using sigmoid functions are capable of approximating any continuous functions on a subset of $R^n$ \citep{cybenko1989approximation}. However, the number of neurons needed to do this increases significantly, which could be computationally infeasible. Adding more layers with fewer perceptrons can achieve same results without affecting the performance of the net \citep{hornik1991approximation}. More than three hidden layers are considered deep neural networks (DNN). DNNs extract information or features combining outputs from perceptrons, but the number of weights and data needed to train them significantly increases \citep{lecun2015deep}.
    
     To train artificial neural networks we find the weights that minimize a loss function. For classification purpose, one of the most use loss functions is the categorical cross-entropy for unbalanced datasets \citep{de2005tutorial}. Initially, weights are chosen at random and are updated between epochs. We compare the desired output with the actual one and pursue to minimize the loss function using backpropagation with any form of Stochastic Gradient Descent (SGD) \citep{ruder2016overview}. Then we update each weight using the inverse of the gradient and a learning rate as shown in \citealt{werbos1990backpropagation}. 
     
     Training artificial neural networks with backpropagation can be slow. Many methods have been proposed based on stochastic gradient descent (SGD) \citep{ruder2016overview}. The massive astronomical datasets make training infeasible in practice, and mini-batches are used to speed up the process \citep{lecun1998efficient}. A training epoch corresponds to a pass over the entire dataset, and usually, many epochs are needed to achieve good results. The way weights are updated can change as well. One of the most widely used optimizers has been Adam optimizer as describe in \citealt{kingma2014adam}. It relies on the first moment (mean) and second moment (variance) of the gradient to update the learning rates. \citealt{ruder2016overview} present an overview of the different gradient descent optimizers and the advantages and disadvantages for each one.
  
  \begin{figure}    \includegraphics[width=\columnwidth]{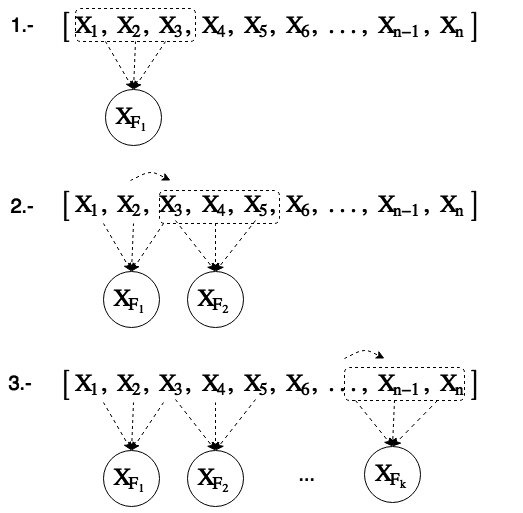}
    \caption{Step by step of a convolution process. A sliding window and a moving step are applied to the data and given as an input to the next layer.}
    \label{fig:convolution}
\end{figure} 
  
    \subsection{Convolutional Neural Nets}
       Convolutional neural networks (CNN) are a type of deep neural network widely used in images \citep{krizhevsky2012imagenet,lecun2015deep}. It consists of an input and output layer as well as several hidden layers different from fully connected ones. 

      A convolutional layer is a particular type of hidden layer used in CNNs. Convolutional layers are in charge of extracting information using a sliding window.
As shown in Figure \ref{fig:convolution} the window obtains local patterns from the input and combines them linearly with its weights (dotted line). Then apply a nonlinear function and pass it to the next layer. The sliding window moves and extracts local information using different inputs but with the same weights. The idea is to specialize this window to extract specific information from local data updating its weights. The size of the window, as well as the moving step, are chosen before running the architecture. Each window corresponds to a specific filter. The number of windows is chosen beforehand. The number of filters can be seen as the number of features we would like to extract. Convolutions are widely used because of their capacity of obtaining features with their translation invariant characteristic using shared weights \citep{krizhevsky2012imagenet}. \citealt{zeiler2014visualizing} study the importance of using filters inside each convolutional layer and show the activation process of using different filters on the Imagenet dataset. 
    
    After applying convolutional layers, a fully connected layer is used to mix the information extracted by the convolutional layers. Fully-connected hidden layers are added to create more complex representations. Finally, the output layer has as many nodes as classes we need.

\section{Method Description}
    \label{sec:method}
    
\begin{figure*}    \includegraphics[width=\textwidth,]{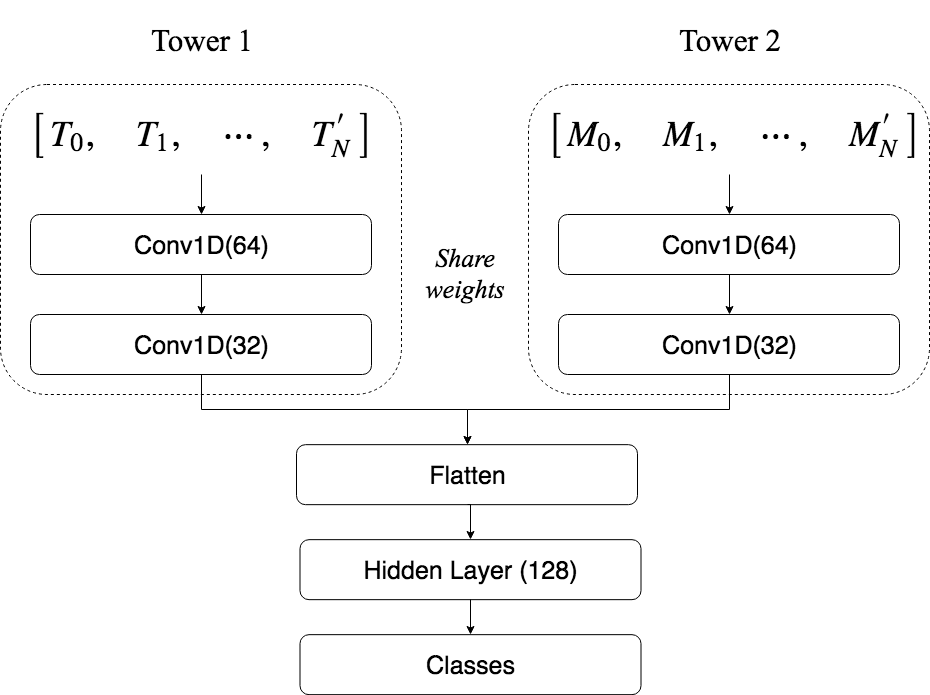}
    \caption{The Convolutional Network Architecture for multi-survey.}
    \label{fig:architecture}
\end{figure*}
    
    We propose an architecture that can classify variable stars using different surveys. We now explain each layer of our architecture, depicted in Figure \ref{fig:architecture}.
    
    Our architecture transforms each light curve to a matrix representation using the difference between points. We use two convolutional layers for extracting the local patterns and turn them into a flat layer. Two fully connected layers are used, and an output layer is plugged at the end to perform the classification. In the following subsections, we describe and give insights on each of the layers.

\subsection{Pre-processing}
    In this phase, light curves are transformed into a matrix. Having a balanced dataset is critical for our purpose of multi-survey classification. Therefore, we use $N_{Max}$ as the maximum number of stars we can extract per class and survey. Section \ref{sec:data_extraction_augmentation} explains in detail the selection of the light curves for the database.
    
\begin{figure}    \includegraphics[width=\columnwidth,]{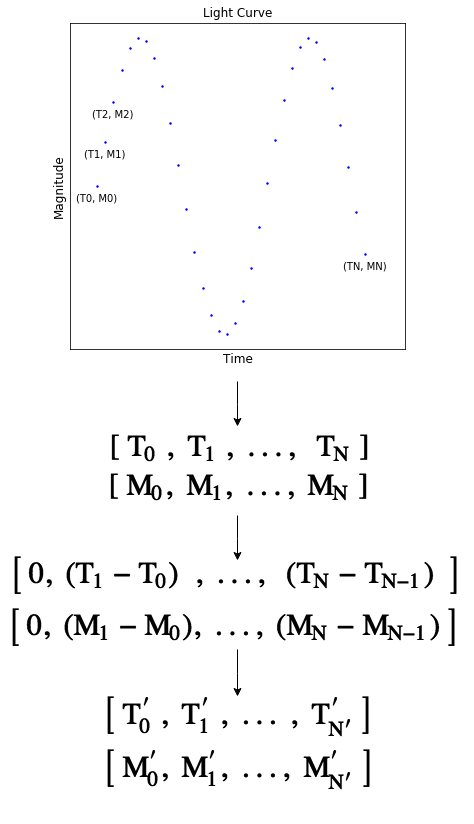}
    \caption{Each light curve is transformed using a matrix with two channels, time and magnitude. For every channel, the maximum number of points is $N$. Each light curve is transformed using the difference between points and a new matrix is created with two channels.}
    \label{fig:matrix_representation}
\end{figure}

    We transform each of these light curves in a matrix representation of size $2 \times N$ where $2$ corresponds to the number of channels (time and magnitude) and $N$ to the number of points used per light curve. Figure \ref{fig:matrix_representation} shows an example of a light curve in a matrix representation. 
    
    To compare light curves between catalogs a reshape to the matrix must be made. Light curves differ in magnitude and time and for comparing them the difference between observations was used. A matrix of size $M \times 2 \times N$ was created where $M$, $2$ and $N$ corresponds to the number of light curves, channels, and numbers of observations used. Figure \ref{fig:matrix_representation} shows an example of the transformation of a light curve. Section \ref{sec:data_extraction_difference} explains in detail this part of the process.

\subsection{First Convolution}
    We apply a convolutional layer to each of the channels in separate branches with shared weights. We use a shared convolutional layer to preserve the objective of integrating datasets with different cadences. Shared layers mean that each of the filters is the same on every tower. The number of filters is given by $S_1$. We chose $64$ filters, to match the number of features presented in \citealt{nun2015fats}.

    Our convolution does not use a max-pooling layer as done in \citealt{jiang2016cryptocurrency}. We avoid max-pooling because of the low cadence of Vista Survey and the detriment of losing information. The step function $s_w$ was set to $2$ or $12$ days (considering OGLE cadence of 6 days as average) and the sliding window $t_w$ was set to $42$ points or $250$ days as done in \citealt{mackenzie2016clustering,valenzuela2017unsupervised}.
    
\subsection{Second Convolution}
    After applying one convolution, we employ another one to mix and create more complex features. \citealt{jiang2016cryptocurrency} showed that using two convolutions achieves better results.

    As the first convolution, the number of filters is given by $S_2$, and it was established to be half of the filters used in the first convolution.
    
\subsection{Flatten Layer}
    After extracting the local patterns, we transform the last convolution into a flatten layer as in \citealt{jiang2016cryptocurrency,zheng2014time}. Our layer combines its patterns afterwards with a hidden layer in a fully connected way.

\subsection{Hidden Layer}
    We use a hidden layer to combine our extracted patterns, and the number of cells is given by $n_{cells}$. After several experiments, we realize that $128$ cells generate the best results. We perform many experiments using $sigmoid$, $relu$ and $tanh$ activating functions. We obtain the best results using $tanh$ activation, as most of the deep learning literature suggests \citep{lecun1998efficient}.
    
\subsection{Softmax Layer}
    In the output layer, there is one node per each of the possible variability classes. We test two different amount of classes: one for $4$ classes of variable stars and the other for $9$ subclasses. We use a $softmax$ function to shrink the output to the $[0,1]$ range. We can interpret the numbers from the output nodes as the probability that the light curve belongs to the class represented by that node. 
  
    Finally, we minimize the average across training using categorical cross entropy. We use categorical cross entropy as our loss function as we obtained best results and the datasets use are unbalanced.

\section{Data}
\label{sec:data}
  We apply our method to variable star classification using three different surveys: "The Optical Gravitational Lensing Experiment" (OGLE) \citep{udalski2004optical}, "The Vista Variable in the Via Lactea" (VVV) \citep{minniti2010vista} and "Convection, Rotation and planetary Transit" (CoRot) \citep{baglin2002corot,borde2003exoplanet}. We select these surveys because of their difference in cadence and filters. In the following subsections, we explain each of these surveys in detail. 
  
 \subsection{OGLE-III}
  \label{subsec:Ogle}
      The Optical Gravitational Lensing Experiment III (OGLE-III)  corresponds to the third phase of the project \citep{udalski2004optical}. Its primary purpose was to detect microlensing events and transiting planets in four fields: the galactic bulge, the large and small Magellanic clouds and the constellation of Carina. 
      
    For our experiment, we use $451,972$ labeled light curves. The cadence is approximately six days and in the experiments is considered our survey with medium cadence. The band used by the survey is infrared and visible. We discard the visible band because of the low number of observations per star compared to the infrared band. The class distribution is shown in Table \ref{tab:ogle_labels}.
    
\begin{table}
  \centering
    \caption{Class distribution of OGLE-III labeled set.}
    \label{tab:ogle_labels}
    \begin{tabular}{ l l r }
        \hline
        Class name & Abbreviation & Num. of Stars \\
        \hline
        Classical Cepheids & CEP & 8031 \\
        RR Lyrae & RRLyr & 44262 \\
        Long Period Variables & LPV & 343816 \\
        Eclipsing Binaries & ECL & 55863 \\
        \hline
    \end{tabular}
    
    \begin{tabular}{ l l r }
        \hline
        Subclass Name & Abbreviation & Num. of Stars \\
        \hline
        First-Overtone 10 \\ Classical Cepheid & CEP10 & 2886 \\
        Fundamental-Mode F \\ Classical Cepheid & CEPF & 4490 \\
        RR Lyrae ab & RRab & 31418 \\
        RR Lyrae c & RRc & 10131 \\
        Mira & Mira & 8561 \\
        Semi-Regular Variables & SRV & 46602 \\
        Small Amplitude Red Giants  & OSARGs & 288653 \\
        Contact Eclipsing Binary & EC & 51729 \\
        Semi-Detached \\ Eclipsing Binary & nonEC & 4134 \\
        \hline
    \end{tabular}
\end{table}    
  
  \subsection{The Vista Variable in the V\'ia L\'actea}
  \label{subsec:VVV}
    The Visible and Infrared Survey Telescope (Vista) started working in February 2010 \citep{minniti2010vista}. Its mission was to map the Milky Way bulge and a disk area of the center of the Galaxy.
    
    To obtain labeled light curves from Vista, we cross-match the Vista catalog with OGLE-III. We found $246,474$ stars in total. The cadence of the observations are approximately every eighteen days and is considered our survey with low cadence. The band used by the survey is mainly $k_{ps}$, and the class distribution of the labeled subset is shown in Table \ref{tab:vvv_labels}.
  
  \begin{table}
    \centering
    \caption{Class distribution of VVV labelled set.}
    \label{tab:vvv_labels}
    \begin{tabular}{ l l r }
        \hline
        Class name & Abbreviation & Num. of Stars \\
        \hline
        Classical Cepheids & CEP & 36 \\
        RR Lyrae & RRLyr & 15228 \\
        Long Period Variables & LPV & 228606 \\
        Eclipsing Binaries & ECL & 2604 \\
        \hline
    \end{tabular}
    
    \begin{tabular}{ l l r }
        \hline
        Subclass Name & Abbreviation & Num. of Stars \\
        \hline
        First-Overtone 10 \\ Classical Cepheid & CEP10 & 5 \\
        Fundamental-Mode F \\ Classical Cepheid & CEPF & 23 \\
        RR Lyrae ab & RRab & 10567 \\
        RR Lyrae c & RRc & 4579 \\
        Mira & Mira & 8445 \\
        Semi-Regular Variables & SRV & 37366 \\
        Small Amplitude Red Giants  & OSARGs & 182795 \\
        Contact Eclipsing Binary & EC & 1818 \\
        Semi-Detached \\ Eclipsing Binary & nonEC & 786 \\
        \hline
    \end{tabular}
  \end{table}  
    
   \subsection{CoRoT}
   \label{subsec:Corot}
   The Convection, Rotation and planetary Transits (CoRoT) is a telescope launched in December 2016 \citep{baglin2002corot,borde2003exoplanet}. Its main purpose is to continuously observe the milky way for periods up to 6 months and search for extrasolar planets using transit photometry. One of the main advantages is the high cadence that can be more than a 100 observations per object per day. 
   
    Because of its early stage, just a few instances have been labeled. For our experiments, we use $1,311$ labeled light curves. The cadence of the observations are approximately every sixty per day and in the experiments is considered our survey with a high cadence. This catalog does not use any specific filter but the observations per object are in red, blue and green bands and for the experiments, we used the white band combining this three. The class distribution is shown in Table \ref{tab:corot_labels} 
  
   \begin{table}
    \centering
    \caption{Class distribution of CoRoT labelled set.}
    \label{tab:corot_labels}
    \begin{tabular}{ l l r }
        \hline
        Class name & Abbreviation & Num. of Stars \\
        \hline
        Classical Cepheids & CEP & 125 \\
        RR Lyrae & RRLyr & 509 \\
        Long Period Variables & LPV & 109 \\
        Eclipsing Binaries & ECL & 568 \\
        \hline
    \end{tabular}
    
    \begin{tabular}{ l l r }
        \hline
        Subclass Name & Abbreviation & Num. of Stars \\
        \hline
        RR Lyrae ab & RRab & 28 \\
        RR Lyrae c & RRc & 481 \\
        \hline
    \end{tabular}
  \end{table}    

\section{Data Pre-processing}
    \label{sec:data_extraction}
    In this section, we explain in detail how to pre-process the data to produce our architecture inputs. Given that we are integrating several surveys that contain light curves with different bands and number of observations, we have to pre-process the data to get a survey-invariant architecture.
 
    \subsection{Time \& Magnitude Difference}
   \label{sec:data_extraction_difference}
   The difference between instruments, cadences and bands between surveys create a bias in the variability patterns of light curves \citep{long2012optimizing, benavente2017automatic}. 
We use as an input the difference between the time and magnitude with the previous measurements. This removes the extinction, distance and survey specific biases. Moreover, it acts as a normalization method. It enables the network to learn patterns directly from the observations without the need to pre-processing any of the data or extinction correction.
 
   \subsection{Light Curve padding}        
    The difference in cadence among OGLE-III, Vista, and Corot catalogs create a big variance in the number of observations per light curve. To overcome this problem, we impose a minimum number of observations and use a zero padding to complete the light curves that cannot reach that minimum. This is inspired by the padding procedure done in deep learning for image analysis. To define such limit, we tried many different values and notice that classification results do not change significantly within a range of $500$ and $1,500$ observations. We fixed the limit at $500$ points per light curve because that amount preserves the classification accuracy and keeps most of the light curves suitable for our analysis. 
   
   \subsection{A Light curve data augmentation model}
    \label{sec:data_extraction_augmentation}

\begin{figure}
\includegraphics[width=\columnwidth,]{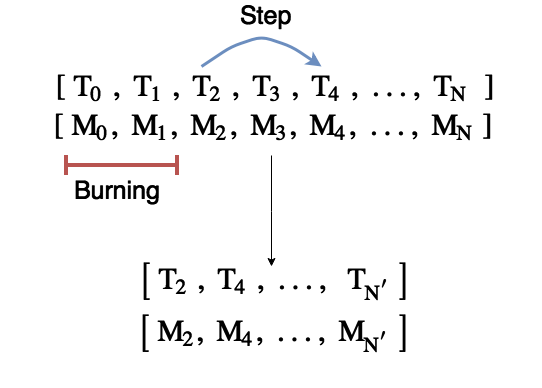}
    \caption{Example of a new light curve using a burning parameter of 2 and a step parameter of 1.}
    \label{fig:data_augmentation}
\end{figure}

    \citealt{hensman2015impact} studied the impact of unbalanced datasets in convolutional neural networks and proved that for better classification performance the dataset should be class-balanced. Since the datasets used in this paper are unbalanced, data augmentation techniques have to be applied \citep{krizhevsky2012imagenet}.

    To balance the dataset, we propose a novel data augmentation technique based on light curve replicates. As mention before in Section \ref{sec:method}, the number of stars per class and survey is given by $N_{Max}$. If the number of light curves per class and survey is larger than this parameter, the replication process does not take place. Otherwise, the light curves are replicated until they reach that limit. Each class is replicated using two light curve parameters: $burning$ and $step$. The $burning$ parameter indicates how many points we have to discard in the light curve. The $step$ parameter tells every how many points we should take samples. The $burning$ parameter goes up every time the light curve has been replicated making the starting point different for each new data of a specific class and survey. Figure \ref{fig:data_augmentation} shows an example of the replication of a light curve. Is important to note for surveys with high and medium cadences such as Corot and OGLE-III, the loss is not critical as many observations are available. In cases of low cadence catalogs, such as Vista, the loss of observations is significantly reduced depending on the $step$ parameter. To keep a minimum observation loss, the $step$ parameter is set to a random number between $0$ and $2$. The maximum replication of a light curve is $5$.

\section{Parameter Initialization}
    \label{sec:par_initialization}
     As in most of deep learning solutions, our architecture needs the set up of several initial parameters. In this section, we explain the model design and how to set up its parameters.
    
    \subsection{Parameters} 
   As previously noted, surveys have different optics and observation strategies which impact the depth at which they can observe. This impacts the number of variable stars detected and cataloged. In our case, OGLE has been operating longer and observes large portions of the sky while VVV goes deeper but in a smaller area, and Corot has great time resolution but is considerably shallower. The combined catalog is dominated by OGLE stars and the subclasses are highly unbalanced, being the LPV class the majority of them.
   
	In order to train with a more balanced dataset, we use a limit of $8,000$ stars per class and survey. We test different values and set it to $8,000$ as most of the classes and subclasses of VISTA and OGLE survey possess that amount as shown in section \ref{sec:data}. 
    Finally, after several experiments measuring the training speed and efficiency, we set the batch size to $256$. Table \ref{tab:parameters} shows a summary of the parameters of our architecture.
    
    \subsection{Layers}
       We use two convolutional layers as done in \citealt{jiang2016cryptocurrency}. In the imaging literature, several works show that one convolutional layer is not enough to learn a suitable representation, and commonly they use two convolutions \citep{zheng2014time,jiang2016cryptocurrency,gieseke2017convolutional}. 

We try using only one convolution and performance was critically reduced. Three convolutions are also utilized, producing results as good as using two, but the time for training the net and the number of parameters increase significantly.
        
        A window size $t_w$ was used for the convolution process and set to $42$ observations or $250$ days in average as done in \citealt{mackenzie2016clustering,valenzuela2017unsupervised}. Finally, a stride value $s_w$ was used and set to $2$ or 12 days in average as done in \citealt{mackenzie2016clustering,valenzuela2017unsupervised}.
        
    \subsection{Activation functions}
        In convolutional layers we used \textit{relu} activation function as they are capable of extracting the important information .
\begin{equation}
relu(x) \quad = \quad \max( 0, x))
\end{equation}
        For hidden layers, except for convolutional, we used \textit{tanh} functions because of better results, there widely used and better gradients they provide \citep{lecun1998efficient}.

       \subsection{Dropout}
      Dropout is a regularization technique which randomly drops units in the training phase \citep{srivastava2014dropout}. \citealt{srivastava2014dropout} prove the importance of dropout in neural networks to prevent overfitting. We used a dropout of $0.5$ as suggested in the mentioned work in the two fully connected parts of our architecture. One between the flattening layer and hidden layer, the other between the hidden layer and the output layer. Dropout increases the generalization of the network, therefore the performance.
      
  \begin{table}
    \centering
    \caption{Parameters used for the proposed architecture}
    \label{tab:parameters}
    \begin{tabular}{ l l r }
        \hline
        Parameter name & Abbreviation & Value \\
        \hline
        Global Parameters \\
        \hline
        Stars per survey and class & $N_{Max}$ & 8000 \\
        Number of Points \\ 
        Per Light Curve & N & $500$\\
        Batch Size & - & 256 \\
        \hline 
        Architecture \\
        \hline
        Filters for first convolution & $S_1$ & 64 \\
        Filters for second convolution & $S_2$ & 32 \\
        Window size & $t_w$ & 42 \\
        Stride Value & $s_w$ & 2 \\
        Perceptrons in the hidden layer & $n_{cells}$ & 128 \\
        Dropout & - & $0.5$ \\
        \hline
        Data Augmentation Parameters \\
        \hline
        Burning & burning & [1,5] \\
        Step & step & [0,2]\\
        \hline
    \end{tabular}
  \end{table}    

\section{Results}
    \label{sec:results}
    
\begin{figure}
\includegraphics[width=\columnwidth,]{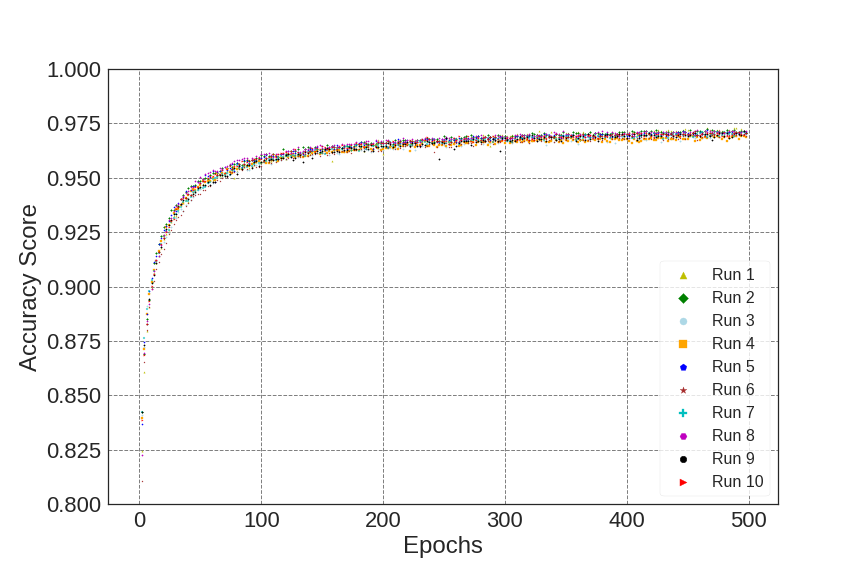}
    \caption{Accuracy of the training model using a 10-fold stratified cross validation with classes.}
    \label{fig:ac_classes}
\end{figure}

\begin{figure}
\includegraphics[width=\columnwidth,]{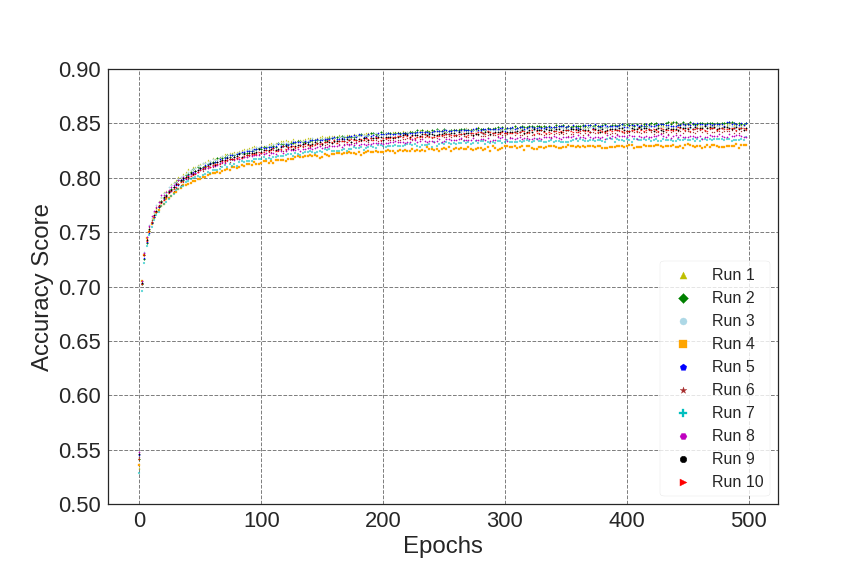}
    \caption{Accuracy of the training model using a 10-fold stratified cross validation with subclasses.}
    \label{fig:ac_subclasses}
\end{figure}
    
    Our experiments are mainly intended to evaluate the classification accuracy of the datasets described in Section \ref{sec:data} as well as the time taken for training the model. We test our model using classes and subclasses and compare it with a Random Forest classifier (RF) \citep{breiman2001random}, still the most used model for light curve classification \citep{richards2011machine,dubath2011random, long2012optimizing,gieseke2017convolutional}. 
In classification accuracy, our model achieves better results in ECL classes and subclasses and comparable results in LPV and RR Lyrae. RF produces better results in OGLE-III dataset.

    Fifty-nine features were extracted for each of the surveys using FATS library, computed over $500$ observations per light curve. For more information about the features extracted please refer to \citealt{nun2015fats}. We use the extracted features to train the RF. 
    
    For all of the experiments, we perform a 10-fold stratified cross-validation. The stratification is done by class and survey to keep the same proportions of classes per survey in the training and testing sets. We perform data augmentation on each of the steps of the cross-validation within the training set. Data augmentation technique is important for improving results in both models. Finally, a $20\%$ validation set is used in our model to control the number of epochs during the training process.

\subsection{Computational Run Time}
\label{sec:time_analysis}

 As mentioned before, the LSST will start working in 2022 generating approximately 15TB per night. The vast amount of data arriving in the future will demand scalable algorithms. We measure the execution time for each of the classification algorithms, considering the feature extraction (needed just for the RF) and the training iterations. In Table \ref{tab:time} we can see that because of the feature-extraction required in RF, the RF model takes way more time in overall. RF is faster than CNN in the training phase because CNN needs several epochs, in this case, $500$ epochs, where each one takes between $6$ and $11$ seconds to run for classes and subclasses respectively. The extraction of features is parallelized using CPUs, and the model training is parallelized using GPUs. Note that a significant improvement in the feature extraction process could be made if FATS library supported GPUs.

\begin{table}
    \centering
    \caption{Approximately time of extraction of features and training the algorithms.}
    \label{tab:time}
    \begin{tabular}{ l r r r }
        \hline
        Method & Extraction & Training & Total  \\
        &  of Features &  Algorithm & Run Time\\
        \hline
        RF & 11.5 days & \bf 36 min &  11.52 days\\
        Classes CNN  & \bf 30 min & 50 min & \bf 1.33 hrs\\
        Subclasses CNN  & \bf 30 min & 91.8 min & \bf 2.03 hrs\\
        \hline
    \end{tabular}
    
    \caption{Accuracy per class and subclass for each survey.}
    \label{tab:accuracy}
    \begin{tabular}{ l c c }
        \hline
        Class & CNN & RF \\
        \hline
		ECL-OGLE & \bf 0.98 $\pm$ 0.01  & 0.97 $\pm$ 0.01  \\
        ECL-VVV & \bf 0.92 $\pm$ 0.02  & 0.89 $\pm$ 0.03  \\
        ECL-Corot & 0.00 $\pm$ 0.00  & \bf 0.91 $\pm$ 0.04	 \\
        LPV-OGLE & \bf 0.99 $\pm$ 0.00  & 0.97 $\pm$ 0.01	 \\
        LPV-VVV & 0.94 $\pm$ 0.01  & \bf 0.97 $\pm$ 0.01	 \\
        LPV-Corot & \bf 0.92 $\pm$ 0.11  & 0.00 $\pm$ 0.00		 \\
        RRLyr-OGLE & 0.94 $\pm$ 0.01  & \bf 0.97 $\pm$ 0.00	 \\
        RRLyr-VVV & \bf 0.94 $\pm$ 0.01  &  0.86 $\pm$ 0.02	 \\
        RRLyr-Corot  & 0.00 $\pm$ 0.00  & \bf 0.58 $\pm$ 0.07		 \\
        CEP-OGLE & 0.90 $\pm$ 0.03  & \bf 0.93 $\pm$ 0.01	 \\
        CEP-VVV & 0.00 $\pm$ 0.00  & \bf 0.08 $\pm$ 0.17		 \\
        CEP-Corot & \bf 0.90 $\pm$ 0.08  & 0.46 $\pm$ 0.12		 \\
        \hline
        Subclass & CNN & RF \\
        \hline
        EC-OGLE &       \bf 0.93 $\pm$ 0.01  & 0.91 $\pm$ 0.01    \\
        EC-VVV &        0.69 $\pm$ 0.04  & \bf 0.71 $\pm$ 0.03              \\
        nonEC-OGLE &    \bf 0.98 $\pm$ 0.01  & 0.94 $\pm$ 0.01    \\
        nonEC-VVV & \bf 0.51 $\pm$ 0.04  & 0.37 $\pm$ 0.05            \\
        Mira-OGLE & \bf 0.98 $\pm$ 0.01  & \bf 0.98 $\pm$ 0.00    \\
        Mira-VVV & \bf  0.94 $\pm$ 0.01  & 0.29 $\pm$ 0.02        \\
        SRV-OGLE &      0.92 $\pm$ 0.02  & \bf 0.93 $\pm$ 0.01    \\
        SRV-VVV &       0.00 $\pm$ 0.00  & \bf 0.67 $\pm$ 0.02    \\
        Osarg-OGLE & \bf 0.90 $\pm$ 0.01  & 0.88 $\pm$ 0.01   \\
        Osarg-VVV &     0.00 $\pm$ 0.00  & \bf 0.72 $\pm$ 0.01    \\
        RRab-OGLE &     0.72 $\pm$ 0.03  & \bf 0.96 $\pm$ 0.01    \\
        RRab-VVV &      0.77 $\pm$ 0.02  & \bf 0.83 $\pm$ 0.01    \\
        RRab-Corot & 0.11 $\pm$ 0.15  & \bf 0.48 $\pm$ 0.29             \\
        RRc-OGLE &      0.86 $\pm$ 0.02  & \bf 0.98 $\pm$ 0.00    \\
        RRc-VVV &       0.75 $\pm$ 0.03  & \bf 0.83 $\pm$ 0.02    \\
        RRc-Corot &     0.01 $\pm$ 0.01  & \bf 0.99 $\pm$ 0.01                \\
        CEP10-OGLE & 0.84 $\pm$ 0.03  & \bf 0.92 $\pm$ 0.02      \\
        CEP10-VVV &     0.00 $\pm$ 0.00  & 0.00 $\pm$ 0.00                  \\
        CEPF-OGLE &     0.72 $\pm$ 0.02  & \bf 0.90 $\pm$ 0.01    \\
        CEPF-VVV &      0.00 $\pm$ 0.00  & 0.00 $\pm$ 0.00              \\
        \hline
    \end{tabular}
\end{table}

 Our proposed architecture and RF are trained using a computer with $128$ GB RAM, a GeForce GTX 1080 Ti GPU and 6 CPU. Our algorithm is developed using Keras \citep{chollet2015keras} framework that runs on top of Tensorflow \citep{tensorflow2015-whitepaper} library. We use the scikit-learn \citep{scikit-learn} implementation for RF with defaults settings except for the minimum samples leaf that was set to a 100 for better accuracy. 

 We can see that our method is significantly faster as it works with raw magnitudes and time and requires only a couple of minutes of pre-processing. 
    
    \subsection{Results with general classes of variability.}
\begin{figure*}
\includegraphics[width=\textwidth,]{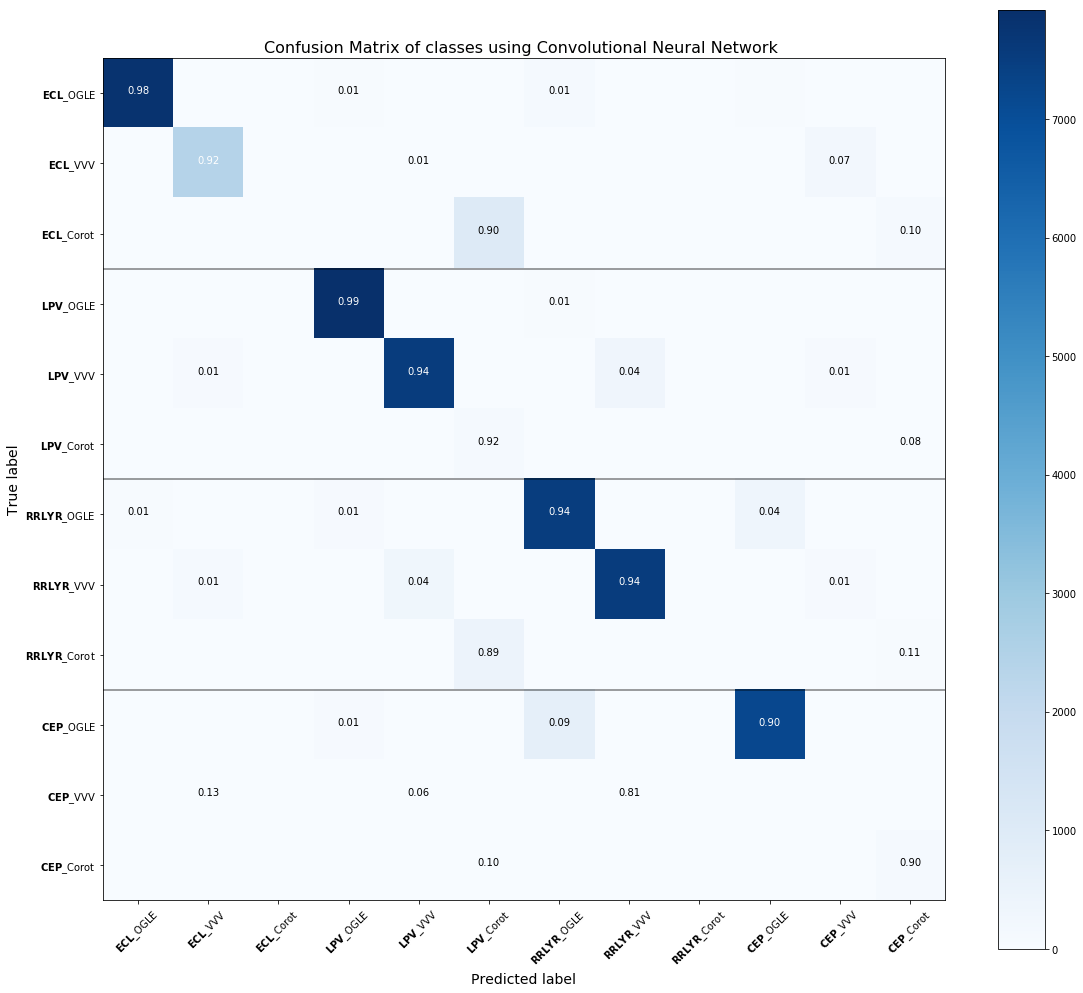}
    \caption{Confusion matrix per class and survey for the convolutional neural network. Empty cells correspond to 0\%.}
    \label{fig:clases_cnn}
\end{figure*}

\begin{figure*}
\includegraphics[width=\textwidth,]{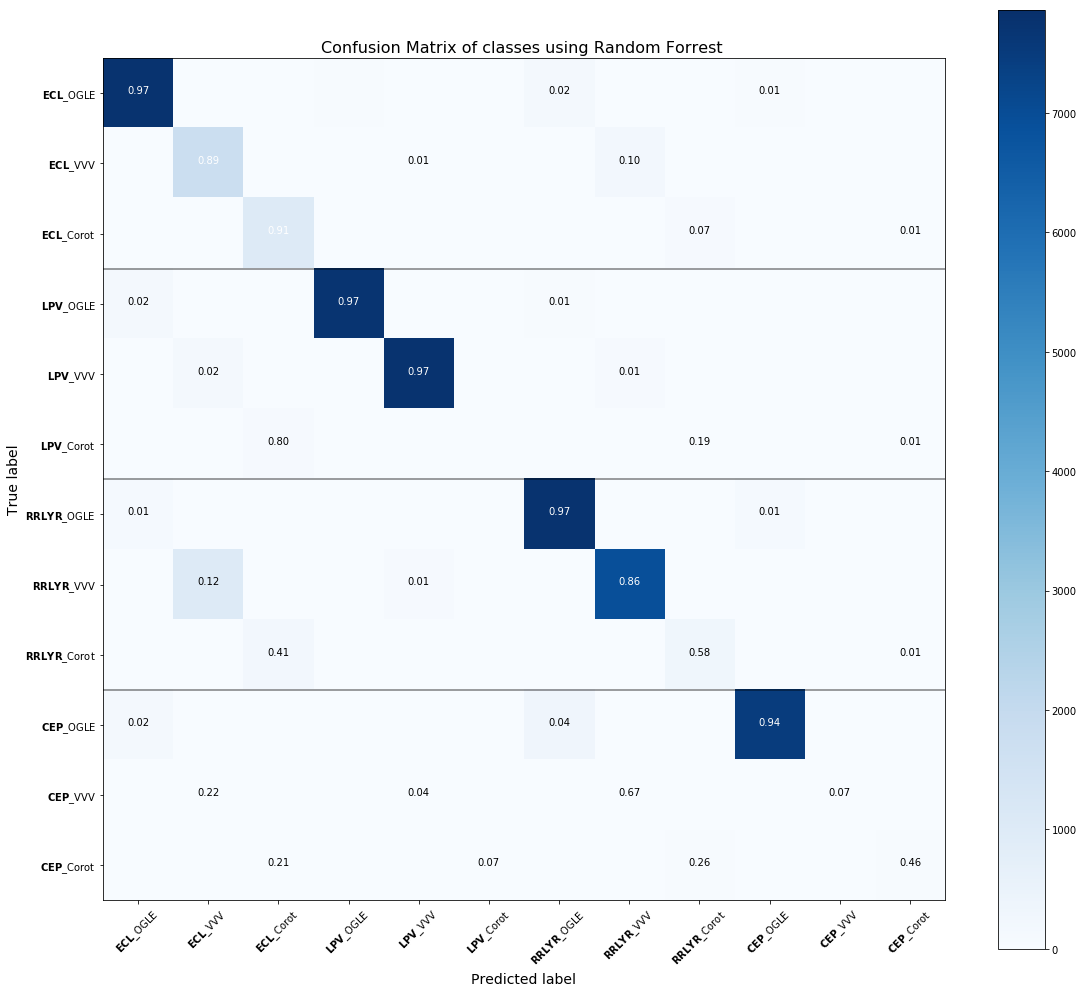}
    \caption{Confusion matrix per class and survey using Random Forest algorithm. Empty cells correspond to 0\%.}
    \label{fig:clases_rf}
\end{figure*}

    We test our model using four general classes: \begin{enumerate*}\item \textit{Cepheids} (CEP), \item \textit{Long Period Variables} (LPV), \item \textit{RR Lyrae} (RRlyr) and \item \textit{Eclipsing Binaries} (ECL). \end{enumerate*} The distribution of classes and subclasses per survey are shown in Tables \ref{tab:ogle_labels}, \ref{tab:vvv_labels} and \ref{tab:corot_labels}. Figure \ref{fig:clases_cnn} and \ref{fig:clases_rf} show the results of using our convolutional architecture and RF respectively. Table \ref{tab:accuracy} summarize the accuracy per class of both approaches.
    
     As it can be seen, RF achieves $96\%$ of accuracy in OGLE-III dataset as it has more labeled data than the others surveys. In VVV RF obtains $97\%$ of accuracy in some of the classes that have more labeled data (\textit{LPV}), but not in stars with few labeled examples (\textit{ECL} and \textit{CEP}). In Corot, RF achieves $91\%$ of accuracy only in \textit{ECL} stars, mainly because of the high cadence of Corot that make it infeasible to extract features correctly, especially those related to periodicity. RF results show that FATS features of some light curves (\textit{LPV} and \textit{RRlyr}) can classify accurately between different surveys. That is not a surprise mainly because period features are less sensitive to changes in cadence.   
    
    Our proposed architecture achieves comparable classification accuracy in OGLE-III but with much less training time.  As shown in Figure \ref{fig:ac_classes} our model produces approximately an accuracy of $97\%$ in the validation set. Each of the colors represents one training of the 10-fold stratified cross-validation. As shown in Table \ref{tab:accuracy}, OGLE-III dataset achieves $95\%$ of accuracy in average in all of its classes. VVV survey achieves $93.7\%$ of accuracy in most of its classes, except for CEP stars, which are less than $40$ light curves. Comparing it to RF performance, it achieves better performance in VVV as it has $92\%$ of accuracy on each of the classes, except for CEP. In Corot, CNN and RF achieve comparable results.  
    
    \subsection{Results with subclasses of variability}
\begin{figure*}
\includegraphics[width=\textwidth,]{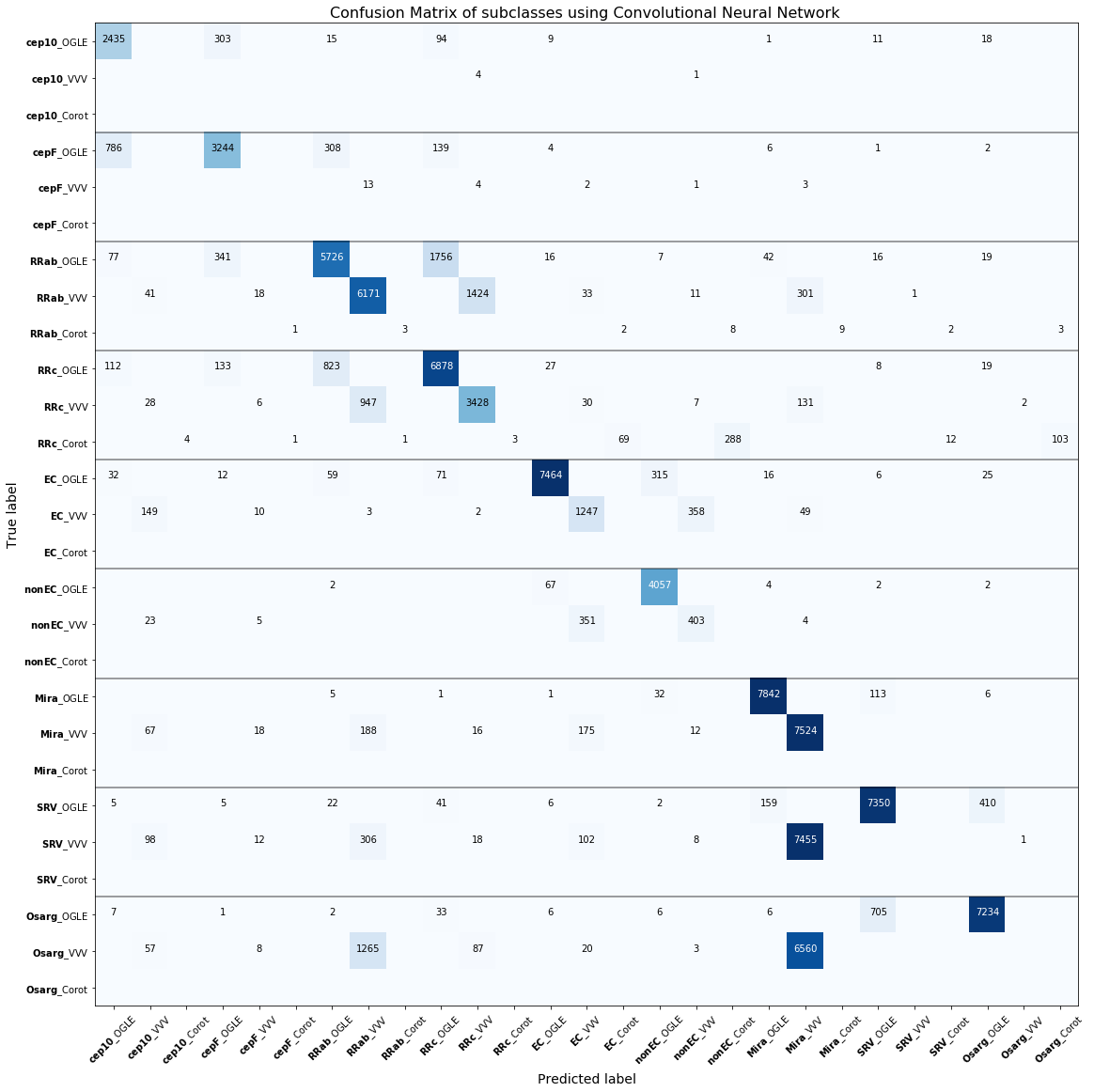}
    \caption{Confusion matrix per class and survey for the convolutional neural network. Empty cells correspond to 0.}
    \label{fig:subclases_cnn}
\end{figure*}

\begin{figure*}
\includegraphics[width=\textwidth,]{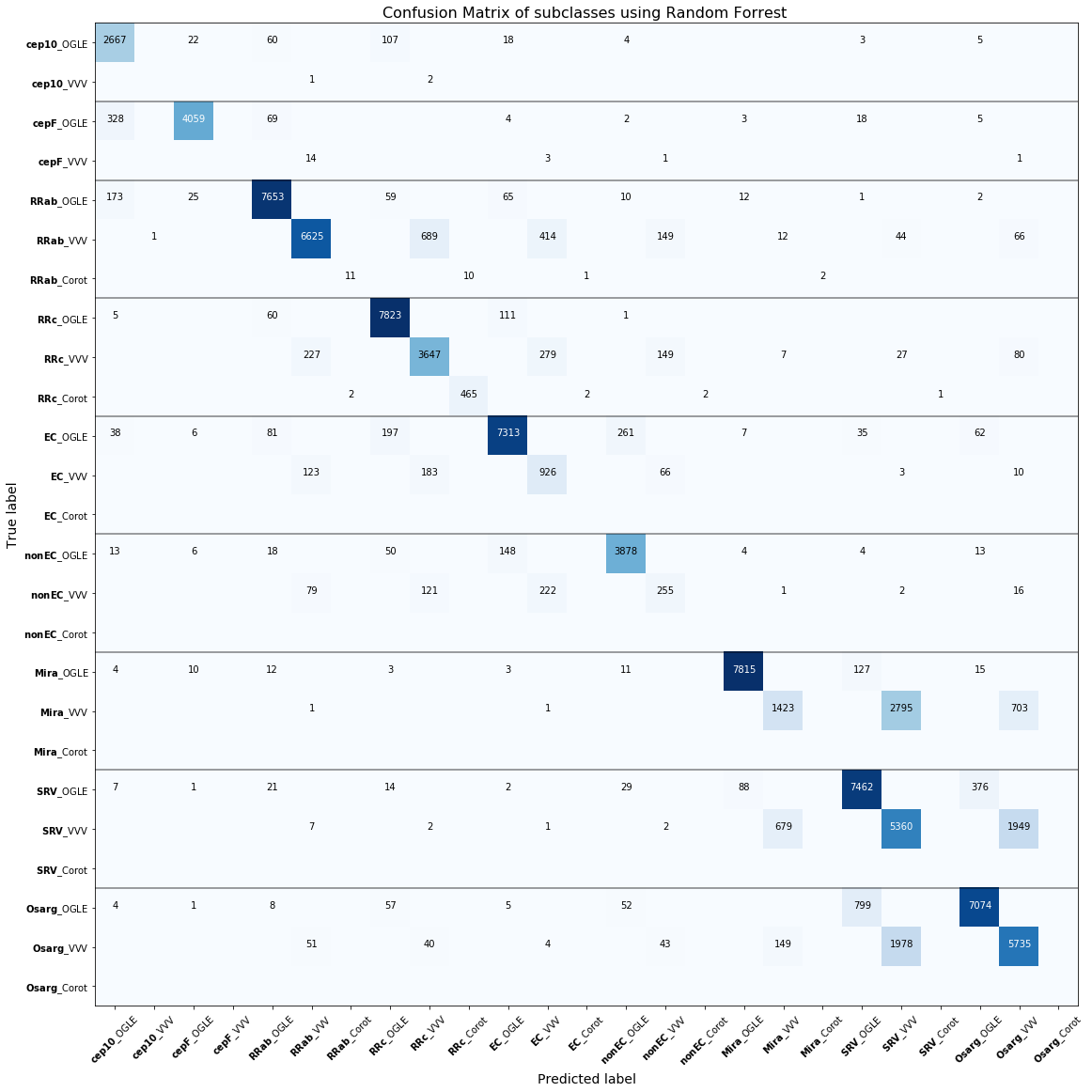}
    \caption{Confusion matrix per class and survey using Random Forest algorithm. Empty cells correspond to 0.}
    \label{fig:subclases_rf}
\end{figure*}

    We test our model using nine types of subclasses: \begin{enumerate*}\item \textit{First-Overtone 10 Classical Cepheid} (Cep10), \item \textit{Fundamental-Mode F Classical Cepheid} (CepF), \item \textit{RR Lyrae ab} (RRab), \item \textit{RR Lyrae c} (RRc), \item \textit{Mira}, \item \textit{Semi-Regular Variables} (SRV), \item \textit{Small Amplitude Red Giants} (OSARGs), \item \textit{Eclipsing Binaries} (EC) and \item \textit{Eclipsing Binaries Semi-Detached} (nonEC). \end{enumerate*} The distribution of subclasses per survey are shown in Tables \ref{tab:ogle_labels}, \ref{tab:vvv_labels} and \ref{tab:corot_labels}. Figure \ref{fig:subclases_cnn} and \ref{fig:subclases_rf} show the results of using our neural network architecture and Random Forest respectively. Table \ref{tab:accuracy} summarize the accuracy per subclass of both approaches.
    
    As shown in Figure \ref{fig:subclases_rf}, RF achieves better accuracy in \textit{RR Lyrae} and \textit{Cepheid} stars, despite the small number of light curves and mainly because of the data augmentation technique. As shown in Table \ref{tab:accuracy}, OGLE-III dataset achieves more than $90\%$ of accuracy in most of the subclasses, as it has more labeled data than the others surveys. However, in VVV survey, a $80\%$ of accuracy is obtained in \textit{RRab} and \textit{RRc} stars. Finally, Corot's catalog achieves $99\%$ of accuracy in \textit{RRc} stars.
      
     As shown in Figure \ref{fig:ac_subclasses}, our model produces approximately an accuracy of $85\%$ in the validation set. In OGLE-III dataset, \textit{nonEC} and \textit{Mira} stars achieve a $98\%$ of accuracy, and $92\%$ in \textit{EC}, \textit{SRV}, and \textit{Osarg} stars. In VVV survey, our model achieves $76\%$ of accuracy in \textit{RRab} and \textit{RRc} stars, and $94\%$ in Mira stars. However, \textit{Osarg} and \textit{SRV} are confused in a $81\%$ and $93\% $ with \textit{Mira} type respectively, which indicates a clear overfitting of LPV stars. None of the models can correctly classify VVV Cepheids, mainly because of the low number of light curves ($28$ light curves in total). With \textit{EC} classes from VVV we achieve better results than RF. Finally, the accuracy achieved in Corot is the lowest, mainly because of the few amounts of light curves used ($28$ \textit{RRab} and $481$ \textit{RRc} stars).  
   
\section{Conclusions}
    \label{sec:conclusion}    
    In this work, we have presented a CNN architecture to classify variable stars, tested on light curves integrated from various surveys. The proposed model can learn from the sequence of differences between magnitude and time, automatically discovering patterns across light curves even with
odd cadences and bands. We show that multi-survey classification is possible with just one architecture and we believe it deserves further attention soon. The proposed model is comparable to RF in classification accuracy but much better in scalability. Also, our approach can correctly classify most of the classes and subclasses of variability. 
  
   Like most of deep learning approaches, our model is capable of learning its light curve representation, allowing astronomers to use the raw time series as inputs. 
   
   In order to have an extra comparison point, we attempt to compare our method with the approach presented in \citet{mahabal2017deep}. We implemented their algorithm; we run it with the same catalogs we use for our method, taking 500 observations per light curve. Mahabal's method did not generate results, because it takes about 5 hrs to process just one light curve. The extra computational cost mainly comes because their algorithm generates the 2D embeddings by comparing every pair of points in each light curve, something that makes it impractical for our experimental setup. To obtain results with Mahabal's approach, we decrease the number of observations per light curve until their method return results in a comparable amount of time. With about 100 observations, their method takes about 5 minutes to run, still too slow for our setup (about 140000 light curves, 8000 light curves per class, etc.) By using ~50 observations, their method could generate results. As we expected, their classification results are much worse, given that the small number of observations make impossible even capture the light curve periods in most of the cases. We are not telling that their method is worse than ours, it is just intended for another problem setup. We believe that it is not a fear comparison to add these results to our paper, because Mahabal's method would be using much less information compared to ours.   
   
   As future work, oversampling techniques should be studied, given that we observe that our model is sensitive to unbalanced training sets. Also, more complex architectures must be developed to improve classification accuracy. New approaches able to deal with few light curves in some classes are needed. For example, simulation models based on astrophysics would be a significant contribution especially for the most unrepresented subclasses of variability. With simulation models, deep learning architectures could be significantly improved, given that usually their performance is directly related to the number of training cases, even with simulated instances. In the same direction, it would be interesting to produce larger training sets by integrating a higher number of surveys. Our code implementation is done in Python, available for download at \url{https://github.com/<authorusername>/DeepMultiSurveyClassificationOfVariableStars}. We also published the catalogs with the cross-matched training sets used in this work. 

\section*{Acknowledgements}

We acknowledge the support from CONICYT-Chile, through the FONDECYT Regular project number 1180054. This paper utilizes public information of the OGLE, Vista and Corot surveys. The computations were made using the cluster from Pontificia Universidad Cat\'olica de Chile. 



\pagebreak
\bibliographystyle{mnras}
\bibliography{bibliography} 







\bsp	
\label{lastpage}
\end{document}